\documentclass[aps,prl,amsmath,amssymb,sort&compress,comma,nofootinbib,floatfix,twocolumn]{revtex4-1}
\usepackage{graphicx}
\usepackage{epstopdf}
\usepackage{subfigure}  
\begin{document}

\newcommand{\etal}{{\it et al.}\/}
\newcommand{\gtwid}{\mathrel{\raise.3ex\hbox{$>$\kern-.75em\lower1ex\hbox{$\sim$}}}}
\newcommand{\ltwid}{\mathrel{\raise.3ex\hbox{$<$\kern-.75em\lower1ex\hbox{$\sim$}}}}

\title{Pairing interaction near a nematic QCP of a 3-band CuO$_2$ model}

\author{T.A. Maier}
\affiliation{Computer Science and Mathematics Division and Center for Nanophase Materials
Sciences, Oak Ridge National Laboratory, Oak Ridge, Tennessee 37831-6494, USA}

\author{D.J.~Scalapino}
\affiliation{Department of Physics, University of California, Santa Barbara, CA 93106-9530, USA}


\begin{abstract}
Here we calculate the strength of the $d$-wave pairing and the $k$ dependence of
the gap function associated with the nematic fluctuations of a CuO$_2$ model as
the doping $p$ approaches a quantum critical point. Higher order $d$-wave harmonics contribute to the $k$ dependence of the resulting superconducting gap function reflecting the longer range nature of the nematic pairing interaction.
\end{abstract}


\maketitle


The importance of a nematic phase and the possible existence of a nematic quantum critical point (QCP) just beyond the optimal doping of the cuprate superconductors was first raised in an article by Kivelson {\it et al}. \cite{Kivelson1}. There are now a variety of experiments \cite{Wu, Fujita,LeBoeuf,Blackburn,Parker} find short-range biaxial charge order in the pseudogap region of the $T$-$p$ ($p$=holes/Cu) phase diagram of the cuprate superconductors.
Ultrasonic measurements \cite{Shekhter} of the elastic moduli of YBaCuO$_{6+\delta}$ crystals
provide thermodynamic evidence of a distinct phase boundary $T^*(p)$ below
which the system is in a pseudo gap phase. Magnetoresistance measurements of the electron effective mass $m^*$ by Ramshaw et al. \cite{Ramshaw} report an increase in $m^*$ as the doping $p$ approaches a critical doping $p_c = 0.18 $ where $T^*(p_c)$  goes to zero. They also find that the magnetic field needed to suppress superconductivity peaks as $p$ approaches $p_c$, clearly implicating the pseudo gap quantum critical fluctuations in the superconducting pairing. There have also been a number of further theoretical ideas regarding the nature of the pseudo gap and its role in superconductivity. Oganesyan {\it et al.} \cite{Kivelson2} discussed the breakdown of Fermi liquid theory at a nematic quantum critical point. Metlitski et al. \cite{Metlitski1, Metlitski2} have argued that near the onset of spin-density wave (SDW) order there is an instability to an Ising nematic charge ordered phase and discussed how nematic critical point fluctuations can mediate pairing. Nie et al. \cite{Nie}  have noted that a number of experimental observations associated with the pseudo gap can be understood if the phase is a nematic remnant of stripe order. Here we examine this problem in the framework of a three band Hubbard model for a CuO$_2$ plane originally introduced by Emery \cite{Emery}.  Bulut et al. \cite{Bulut} have shown, within an RPA approximation, that this model exhibits both SDW and nematic instabilities. We are interested in determining the strength of the $d$-wave pairing  associated with  nematic fluctuations and the $k$ dependence of the gap function as the doping $p$ approaches $p_c$.

In the three-band Hubbard model \cite{Emery} for the CuO$_2$ layer, the single-particle electron creation operators carry an orbital index $\ell$. This index $\ell=1$, 2 or 3 and denotes respectively the $d_{x^2-y^2}$ orbit of the Cu, the $p_x$ orbit of the O$_x$ oxygen and the $p_y$ orbit of the O$_y$ oxygen in the unit cell. The Hamiltonian is
\begin{equation} \label{eq:HCuO}
	H = H_0 + V
\end{equation}
with 
\begin{equation}\label{eq:H0}
	H_0 = \sum_{k,\ell_1,\ell_2,\sigma} (\varepsilon_{\ell_1\ell_2}(k)-\mu) c_{\ell_1\sigma}^\dagger(k)c^{\phantom\dagger}_{\ell_2\sigma}(k)\,.
\end{equation}

\begin{figure}
\includegraphics[width=8cm]{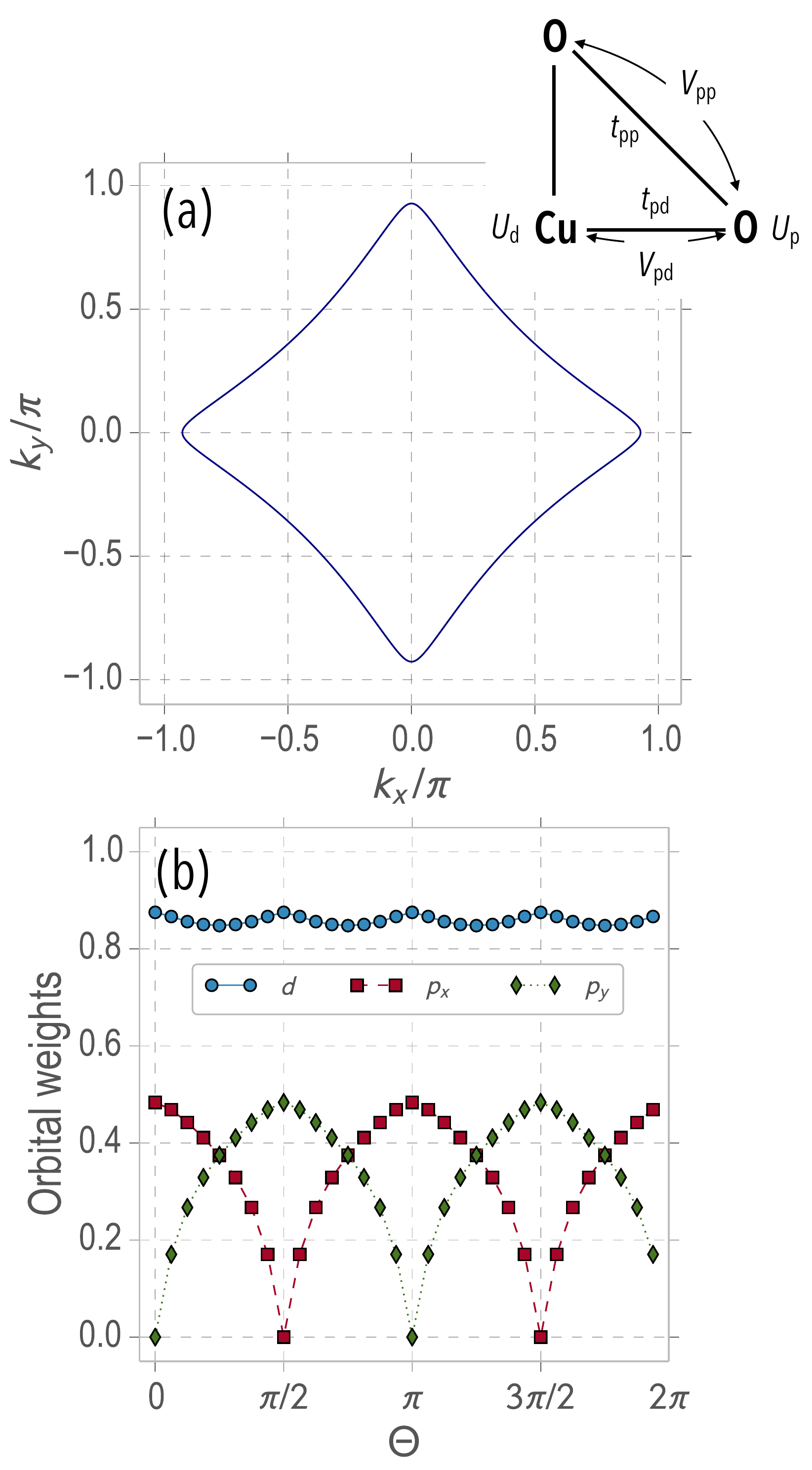}
\caption{(a) The Fermi surface for $ t_{pd}= 1.0 , t_{pp}= 0.5 $ at a doping  $p=0.20.$ The inset shows the hopping and Coulomb interactions parameters.  (b) The orbital weights $|a^l_{\nu}(k)|$  on the Fermi surface. }
\end{figure}

For the hopping parameters shown in Fig.~1a	
\begin{eqnarray}
	\varepsilon_{\ell_1\ell_2}(k) &=& \varepsilon_d \,\delta_{\ell_1,1}\delta_{\ell_2,1}+\varepsilon_p\, [\delta_{\ell_1,2}\delta_{\ell_2,2} + \delta_{\ell_1,3}\delta_{\ell_2,3}]  \\
	&+& 2t_{pd}\sin(k_x/2)\, [\delta_{\ell_1,1}\delta_{\ell_2,2} + \delta_{\ell_1,2}\delta_{\ell_2,1}]\nonumber\\
	&-& 2t_{pd}\sin(k_y/2)\, [\delta_{\ell_1,1}\delta_{\ell_2,3} + \delta_{\ell_1,3}\delta_{\ell_2,1}]\nonumber\\
	&-& 4t_{pp}\sin(k_x/2)\sin(k_y/2)\, [\delta_{\ell_1,2}\delta_{\ell_2,3} + \delta_{\ell_1,3}\delta_{\ell_2,2}]\,.\nonumber
\end{eqnarray}
Diagonalization of $H_0$ gives 3 bands $\nu$ with energy dispersion $E_\nu(k)$.
$V$ contains the Coulomb interactions indicated in Fig.1a and is given by
\begin{equation}
	V = \sum_{\ell_1\ell_2} V_{\ell_1\ell_2}(q) n_{\ell_1}(q)n_{\ell_2}(-q)
\end{equation}
with 
\begin{equation}
	n_{\ell}(q) = \sum_{k,\sigma} c^\dagger_{\ell\sigma}(k+q)c^{\phantom\dagger}_{\ell\sigma}(k)
\end{equation}
and
\begin{eqnarray}
	V_{\ell_1\ell_2}(q) &=& U_d[\delta_{\ell_1,1}\delta_{\ell_2,1}]+U_p[\delta_{\ell_1,2}\delta_{\ell_2,2}+\delta_{\ell_1,3}\delta_{\ell_2,3}]\\
	&+& 2V_{pd}\cos(q_x/2) [\delta_{\ell_1,1}\delta_{\ell_2,2}+\delta_{\ell_1,2}\delta_{\ell_2,1}]\nonumber\\
	&+& 2V_{pd}\cos(q_y/2) [\delta_{\ell_1,1}\delta_{\ell_2,3}+\delta_{\ell_1,3}\delta_{\ell_2,1}]\nonumber\\
	&+& 4V_{pp}\cos(q_x/2)\cos(q_y/2) [\delta_{\ell_1,2}\delta_{\ell_2,3}+\delta_{\ell_1,3}\delta_{\ell_2,2}]\,.\nonumber
	\end{eqnarray}

The effective interaction in the charge channel consists of the direct and the exchange terms
\begin{equation}
	V^c_{\ell_1\ell_2\ell_3\ell_4} = -\delta_{\ell_1\ell_3}\delta_{\ell_2\ell_4} V_{\ell_1\ell_2}(k-k') + \delta_{\ell_1\ell_2}\delta_{\ell_3\ell_4} 2V_{\ell_1\ell_3}(q)\,.
\end{equation}
Within an RPA approximation the charge vertex $\Gamma^c$ is obtained as the solution of the particle-hole t-matrix which sums multiple $V^c$ scatterings. This integral equation can be simplified by writing the interaction in a separable form \cite{Littlewood}
\begin{eqnarray} \label{eq:sepForm}
V_{\ell_1\ell_2}(k-k') &=& \sum_{ij} g^i_{\ell_1\ell_2}(k) \tilde{V}^{ij}_X g^j_{\ell_1\ell_2} (k')\nonumber\\
V_{\ell_1\ell_2}(q) &=& \sum_{ij} g^i_{\ell_1\ell_1} \tilde{V}^{ij}_D(q) g^j_{\ell_2\ell_2}
\end{eqnarray}
The functions $g^i_{\ell_1\ell_2}(k)$ form a 19-dimensional basis.  This basis along with  the 19$\times$19 exchange $\tilde{V}_X$ and direct $\tilde{V}_D$  interaction  matrices  are given in  Appendix A of  Bulut et al. \cite{Bulut}.

The charge vertex is then given by 
\begin{equation}
	\Gamma^c_{\ell_1\ell_2\ell_3\ell_4}(k,k',q) = \sum_{ij} g^i_{\ell_1\ell_2}(k) \tilde{\Gamma}^{ij}_c(q) g^j_{\ell_3\ell_4} (k')	
\end{equation}
with
\begin{equation}
	\tilde{\Gamma}_c(q) = \left[ 1+\tilde{V}_c(q)\tilde{\chi}_0(q) \right]^{-1}\tilde{V}_c(q)\,.
\end{equation}
Here $\tilde{V}_c(q) = 2\tilde{V}_D(q)-\tilde{V}_X(q)$ and
\begin{eqnarray}
\tilde{\chi}^{ij}_0(q) &=& -\frac{1}{N}\sum_{k,\mu,\nu}\sum_{\ell_1,\ell_2,\ell_3,\ell_4} g^i_{\ell_4\ell_3}(k)M^{\ell_1\ell_2\ell_3\ell_4}_{\mu\nu}(k,q)g^j_{\ell_2\ell_1}(k)\nonumber\\
&&\times \frac{f(E_\nu(k+q))-f(E_\mu(k))}{E_\nu(k+q)-E_\mu(k)}
\end{eqnarray}
with $M^{\ell_1\ell_2\ell_3\ell_4}_{\mu\nu}(k,q) = a^{\ell_4}_\mu(k)a^{\ell_2^*}_\mu(k)a^{\ell_1}_\nu(k+q)a^{\ell_3^*}_\nu(k+q)$, where $a^\ell_\nu(k) = \langle c_\ell| \nu k\rangle$ are orbital-band matrix-elements shown in Fig1b.
Within an RPA approximation, the charge susceptibility matrix
\begin{equation}
	\chi_{\ell_1\ell_2}(q) = \int^\beta_0 d\tau \langle {\cal T} n_{\ell_1}(q,\tau) n_{\ell_2}(-q,0) \rangle
\end{equation}
is obtained from $\tilde{\Gamma}_c$ as
\begin{equation}
	\chi_{\ell_1\ell_2}(q) = \chi^0_{\ell_1\ell_2}(q) - \sum_{ij} A^i_{\ell_1\ell_1}(q) \tilde{\Gamma}_c^{ij}(q)A^j_{\ell_2\ell_2}(q)
\end{equation}
with
\begin{eqnarray}
\chi^0_{\ell_1\ell_2}(q) &=& -\frac{1}{N}\sum_{k,\mu,\nu} M^{\ell_1\ell_1\ell_2\ell_2}_{\mu\nu}(k,q) \frac{f(E_\nu(k+q))-f(E_\mu(k))}{E_\nu(k+q)-E_\mu(k)}\nonumber\\
A^i_{\ell_3\ell_4}(q) &=& \frac{1}{N}\sum_{k,\mu,\nu} \sum_{\ell_1,\ell_2} M^{\ell_1\ell_2\ell_3\ell_4}_{\mu\nu}(k,q)g^i_{\ell_1\ell_2}(k)\nonumber\\
&&\hspace{1.5 cm} \times \frac{f(E_\nu(k+q))-f(E_\mu(k))}{E_\nu(k+q)-E_\mu(k)}
\end{eqnarray}
The $d$-wave nematic susceptibility is given by the d-wave projection of the charge susceptibility
\begin{equation}\label{eq:chiN}
	\chi_N(q) = \chi_{xx}(q)+\chi_{yy}(q)-\chi_{xy}(q)-\chi_{yx}(q)\,.
\end{equation}

The charge vertex enters the pairing channel as illustrated in the inset of Fig.~3. A measure of the pairing strength and the $k$-dependent structure of the gap function \cite{Graser} are given by the leading eigenvalue and eigenfunction of
\begin{equation}\label{eq:pairing}
	\oint \frac{dk'_\parallel}{2\pi v_F(k'_\parallel)} \Gamma_c(k,k')\phi_\alpha(k') = \lambda_\alpha \phi_\alpha(k)
\end{equation}
with
\begin{eqnarray}\label{eq:Gammac}
	\Gamma_c(k,k') = &&\sum_{\ell_1,\ell_2,\ell_3,\ell_4} a^{\ell_2^*}_\nu(k')a^{\ell_3^*}_\nu(-k')\Gamma^c_{\ell_1\ell_2\ell_3\ell_4}(k,-k',k-k')\nonumber\\
	&& \hspace{1.5cm}\times a^{\ell_1}_\nu(k)a^{\ell_4}_\nu(-k)
\end{eqnarray}
Here the $k_\parallel$ integral in Eq.~(\ref{eq:Gammac}) is over the Fermi surface, $\nu$ is the band index at the Fermi energy and $v_F(k_\parallel)$ the Fermi velocity.
In the following we will measure energy in units of $t_{pd}$ and set $t_{pp}=0.5, ~ \varepsilon_d-\varepsilon_p=2.5, ~U_d=9,~U_p=3, ~V_{pd}=1$ and $V_{pp}=2$. For these parameters  there is a phase transition to a commensurate $q=0$ nematic phase when the doping decreases below a critical doping $p_c\approx 0.20$ \cite{Bulut}.
The Fermi surface for this doping is shown in Fig.~1a. Although the SDW is the leading RPA instability for these model parameters, we will focus on the pairing which arises from the charge channel.  There, we will find that the nematic fluctuations provide the dominant contribution to the $d$-wave pairing.
In this RPA formulation we  assume that the energy scale of the nematic fluctuations which drive the pairing is larger than $T_c$ and evaluate the charge pairing vertex $\Gamma_c $ and the nematic  susceptibility $\chi_N$  in the $T\rightarrow 0$ limit. This procedure of course breaks down in the critical regime when $p$ is close to $p_c$.  Pairing near an Ising-nematic QCP has been discussed in \cite{Metlitski2}. 

\begin{figure}
\includegraphics[width=8cm]{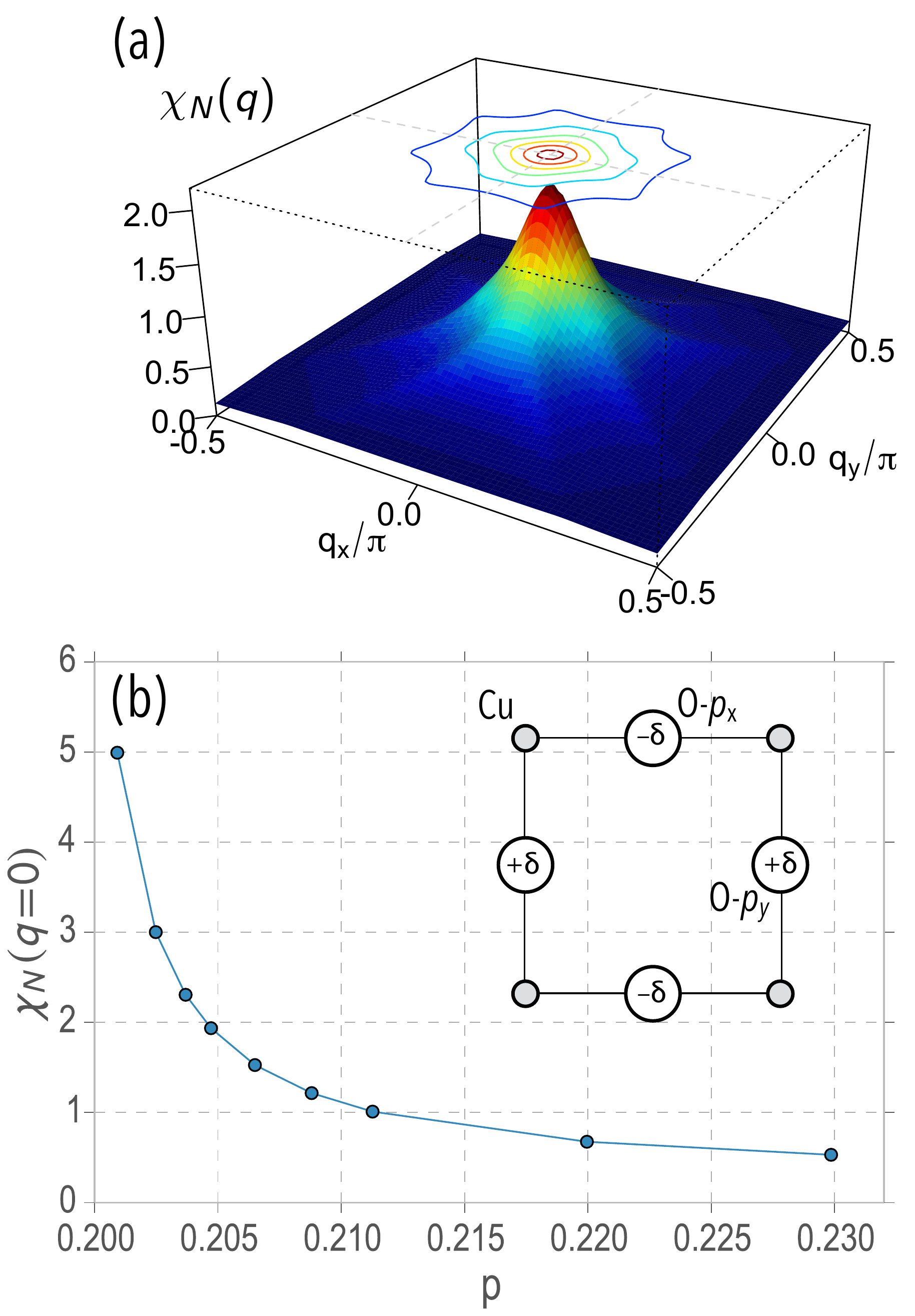}
\caption{The RPA $d$-wave nematic charge susceptibility $\chi_{N}(q)$ verses $q$ for $p=0.205$. (b) $\chi_N(q=0) $ versus $p$. The inset illustrates the nematic mode charge fluctuation on the oxygens.}
\end{figure}

The  nematic ($d$-wave) charge susceptibility $\chi_N(q)$  evaluated from  Eq.~(\ref{eq:chiN}) is shown in Fig.~2a for a doping $p=0.205$. As discussed by Bulut {\it et al.} \cite{Bulut}, depending upon the model parameters, commensurate $q=(0,0)$ , diagonal $q=(q_0,q_0)$ or Cu-O-Cu bond aligned $q=(q_0 ,0)$, $ (0,q_0)$ phases can occur. In all these cases the charge transfer is dominantly between the $O-p_x$ and $O-p_y$ oxygen orbitals, as illustrated in the inset of Fig.~2b. This same intra-unit cell breaking of the point group symmetry of the CuO$_2$ lattice was found in a strong coupling limit of the 3-band Emery model \cite{Kivelson3}. While the $q=(q_0,0)$ phase is observed in the cuprates \cite{Blackburn}. For the interaction parameters we have used, the susceptibility diverges at $p_c\approx 0.20$ where there is a commensurate $q=(0,0)$ quantum critical point. As shown in Fig.~2b, the $q=0$ nematic susceptibility rises rapidly as $p$ approaches $p_c$. For a fixed value of $p-p_c$, the strength of the nematic fluctuations depend on the oxygen-oxygen interaction $V_{pp}$ and increase as $V_{pp}$ increases. 

\begin{figure}
\includegraphics[width=8cm]{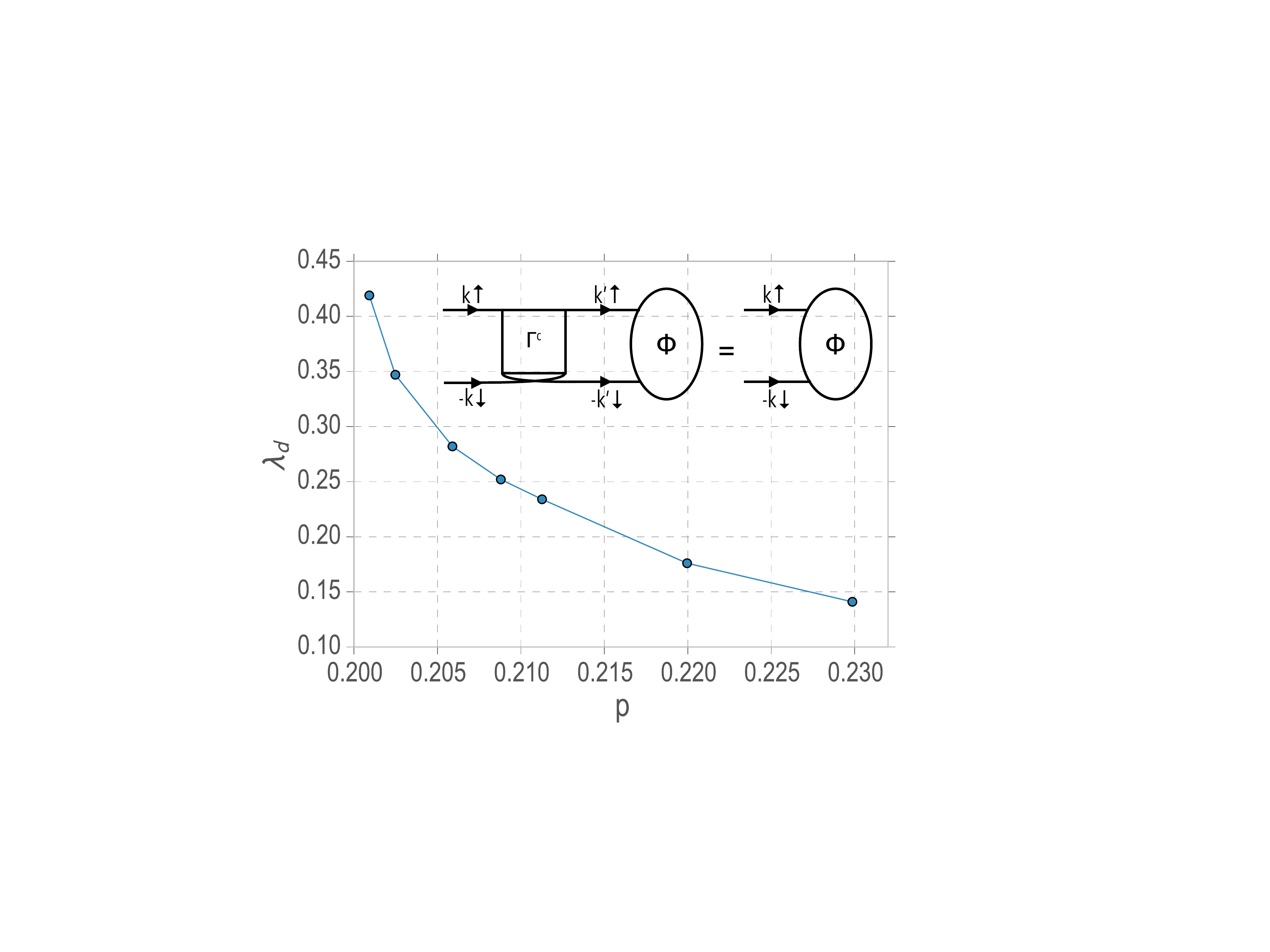}
\caption{The d-wave pairing strength $\lambda_d $ associated with the exchange of  charge fluctuations versus the hole doping $p$. The inset shows a diagrammatic representation of the contribution of the charge vertex $\Gamma_c$  to the gap equation.}
\end{figure}

We turn next to the pairing channel. The eigenfunction of the leading eigenvalue in the pairing channel, Eq.~(\ref{eq:pairing}), has $d$-wave symmetry as seen in Fig.~4. As shown in Fig.~3, the eigen value $\lambda_d$ which is a measure of the $d$-wave pairing strength increases as $p$ approaches $p_c$.  The nematic fluctuations are attractive and contribute positively to the $d$-wave pairing because they involve small momentum transfers.  At a fixed $p-p_c$ value, $\lambda_d$ increases when $V_{pp}$ is increased and the nematic fluctuations increase in strength. In addition, in this RPA treatment, $\lambda_d$ has a strong dependence on the Coulomb interaction $V_{pd}$ between the Cu and the O. While the strength of the nematic fluctuations primarily depend upon $p$ and $V_{pp}$, their contribution to the pairing vertex $\Gamma_c(k,k')$ depends  on $V_{pd}$. This is because $V_{pd}$ provides a coupling of the $O-p_x$ and the $O-p_y$ charge fluctuations to $\Gamma^c_{dddd}$ which as seen in Eq.~(\ref{eq:Gammac}) contributes to $\Gamma^c(k,k')$ through four $d$ orbital weight factors.  In the absence of  the $V_{pd}$ coupling, a coupling of the nematic oxygen charge fluctuations to the $d$-wave pairing channel  involves the product of four $p$ orbital weight factors, which, as seen from Fig.~1b,  is significantly smaller than the product of four $d$-orbital weight factors. Here it is important to remember that the RPA is a weak coupling theory and in strong coupling a large $U_d$ splits the band at the Fermi energy into a lower and an upper Hubbard band. Hole-doping then moves the chemical potential into the band that has predominantly $p$-character and in this case, $V_{pd}$ will play a less important role in the coupling to the nematic fluctuations. 

\begin{figure}
\includegraphics[width=8cm]{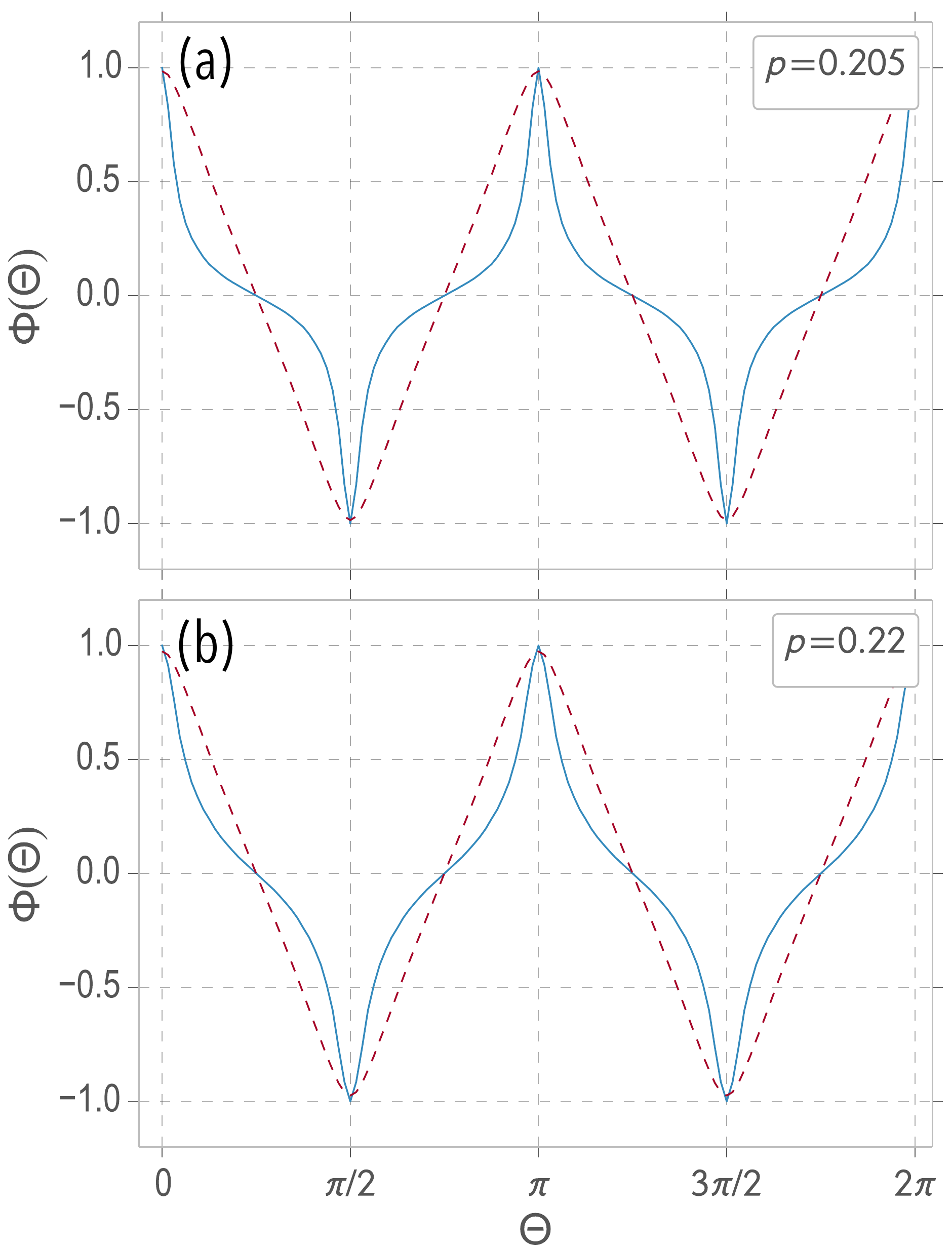}
\caption{The gap function $ \phi_d(k)$ (solid)  versus $\theta$ for $k$ on the Fermi surface normalized to it's value at $\theta=0$ . The dashed curve is $\cos k_x-\cos k_y$. (a) $p=0.205$ and (b) $p=0.22$}
\end{figure}

The gap functions $\phi_d(k)$ for dopings $p=0.205$ and $0.22$ are shown as the solid curves in Figs.~4a and b. The deviation of $\phi_d(k)$ from the $(\cos k_x-\cos k_y)$ form shown as the dashed curves implies the presence of additional higher order $d$-wave harmonics. As one knows, the  pairing interaction associated with  short range  anti-ferromagnetic spin fluctuations  primarily involves near neighbor Cu sites and leads to the familiar $ (\cos k_x-\cos k_y)$ dependence of $\phi_d(k)$.  However, as seen in the plot of $\chi_N(q)$, Fig.~2a, the nematic fluctuations involve small momentum transfers leading to a longer range interaction and increasing the weight of higher harmonics in $\phi_d(k)$. This becomes particularly apparent in the structure of $\phi_d(k)$ as $p$ approaches $p_c$. As noted, YBCO has  Cu-O-Cu bond aligned incommensurate $q^*=(q_0,0)$  or  $(0,q_0)$ nematic fluctuations. In this case, there will be eight regions associated with the Fermi surface points connected by $q^*$ where  $\phi_d(k) $ will exhibit additional structure.  Again, this structure will narrow and peak as p approaches $p_c$. 

Using an RPA approximation for a 3-band model of CuO$_2$ we have shown that nematic charge fluctuations can contribute to the d-wave pairing interaction and that the strength of this pairing increases as the doping $p$ approaches the nematic QCP. The pairing interaction mediated by the nematic fluctuations is ``attractive'' for frequencies below a characteristic fluctuation scale and involves small momentum transfers. Thus it predominantly scatters pairs between ($k$,$-k$) and ($k'$,$-k'$) states on the Fermi surface where the gap has the same sign. Because of the large on-site Cu Coulomb interaction, the pairing occurs in the $d$-wave channel. The small momentum scattering associated with the nematic pairing interaction leads to higher $d$-wave harmonics in the $k$ dependence of the gap, reflecting the longer range nature of the nematic pairing interaction. The nematic fluctuations can work in tandem with the "repulsive" large momentum transfer spin fluctuation interaction so that both the charge and spin  channels contribute to the $d$-wave  pairing strength. In the present weak coupling theory with the bare interaction parameters that we have chosen, an SDW phase transition would onset at a higher temperature precluding the nematic instability. Here we have ignored this and used normal state Green's functions in calculating the nematic susceptibility and the charge vertex. A more complete theory would have different interaction parameters in the spin and charge channels and would require an approach capable of treating strong coupling effects.

\section*{Acknowledgments}

We want to thank W. Hardy, S. Kivelson  and M. Metlitski for useful discussions and acknowledge the support of the Center for Nanophase Materials Sciences, which is sponsored at Oak Ridge National Laboratory by the Scientific User Facilities Division, Office of Basic Energy Sciences, U.S. Department of Energy. As this paper was being completed, we became aware of complementary work by S. Lederer, Y. Schattner, E. Berg, and S. A. Kivelson who studied the problem of enhancement of superconductivity near a nematic quantum critical point. Their conclusions are similar to ours \cite{Kivelson4}.



\end{document}